\documentclass[5p,twocolumn,11pt]{elsarticle}

\usepackage{amsmath,amsfonts,amssymb,amsthm, booktabs, tabularx, placeins}

\usepackage{comment}
\usepackage{physics}
\usepackage{graphicx}
\usepackage[mathscr]{eucal}

\usepackage{subfig}

\newcommand{\be}{\begin{equation}}
\newcommand{\ee}{\end{equation}}
\newcommand{\bea}{\begin{eqnarray}}
\newcommand{\eea}{\end{eqnarray}}
\newcommand{\nn}{\nonumber}

\newcommand{\MP}{M_\text{P}}

\begin{document}

\begin{frontmatter}

\title{Palatini $F(R,X)$: a new framework for inflationary attractors}

\author[Taltech,KBFI]{Christian Dioguardi}

\author[KBFI]{Antonio Racioppi}

\affiliation[Taltech]{organization={Tallinn University of Technology},
            addressline={Akadeemia tee 23}, 
            city={Tallinn},
            postcode={12618}, 
            country={Estonia}
}
\affiliation[KBFI]{organization={National Institute of Chemical Physics and Biophysics},
            addressline={Rävala 10}, 
            city={Tallinn},
            postcode={10143}, 
            country={Estonia}
}

\begin{abstract}
Palatini $F(R)$ gravity proved to be a powerful tool in order to realize asymptotically flat inflaton potentials. Unfortunately, it also inevitably implies higher-order inflaton kinetic terms in the Einstein frame that might jeopardize the evolution of the system out of the slow-roll regime. We prove that a $F(R+X)$ gravity, where $X$ is the inflaton kinetic term, solves the issue. Moreover, when $F$ is a quadratic function such a choice easily leads to a new class of inflationary attractors, fractional attractors, that generalizes the already well-known polynomial $\alpha$-attractors. 
\end{abstract}


\end{frontmatter}

%-------------------------------------------------------------------------------
\section{Introduction} % (fold)
\label{sec:Introduction}

The observation of the cosmic microwave background radiation (CMB) supports the cosmological principle. In order to explain the observed flatness and homogeneity a period of accelerated expansion (inflation) is required in the very early universe \cite{Starobinsky:1980te,Guth:1980zm,Linde:1981mu,Albrecht:1982wi}. With the inflationary mechanism we also achieve to produce primordial inhomogeneities that seed the current large-scale structure of the universe. The most simple model requires a scalar field embedded in Einstein gravity which drives the initial expansion. However, more sophisticated models have shown to achieve interesting results in predicting the CMB observables (e.g. \cite{Martin:2013tda} and refs. therein). The Palatini formulation of gravity (e.g. \cite{Koivisto:2005yc,Bauer:2008zj,Gialamas:2023flv} and refs. therein), in which the Levi-Civita connection is considered to be independent from the metric, shows many appealing features. In particular $F(R)$ models, for which the gravity sector is taken to be a general function of the Ricci scalar, generate asymptotically flat potentials that can be used to describe experimentally viable slow-roll inflation (e.g. \cite{Enckell:2018hmo,Dioguardi:2021fmr,Dioguardi:2022oqu} and refs. therein). However, models that diverge faster than $R^2$, are not well behaved beyond slow-roll \cite{Dioguardi:2022oqu}, hence they could only be used as effective theories during the slow-roll phase. In this paper we extend the class of $F(R)$ theories to the class of $F(R,X)$ where $X$ is the inflaton kinetic term. Models with non minimal couplings involving $R$ and $X$ have been already explored in the past (e.g. \cite{Amendola_1993,Capozziello:1999xt} and refs. therein) but in the metric formulation of gravity. According to our knowledge, this is the first time where such kind of study is performed in the Palatini approach.

In particular we focus on the quadratic models $F(R_X) = 2\Lambda + \omega R_X + \alpha R_X^2$ where $R_X = R + X$. We classify those models and derive their general predictions for inflation introducing two new inflationary attractors that we called $canonical$ and $tailed$ fractional attractors.

%-------------------------------------------------------------------------------

\section{Palatini $F(R,X)$} \label{sec:FRX}
Consider the action
\be
S = \int d^4x \sqrt{-g^J}\qty(\frac{1}{2}F(R_X) - V(\phi)) \label{eq:action:FRX}
\ee
where we assumed Planck units, $\MP=1$, and a space-like metric signature. $V(\phi)$ is the inflaton scalar potential and $F(R_X)$ is an arbitrary function of its argument. We define\footnote{In a similar way, this setup can be easily generalized to $S = \frac{1}{2} \int d^4x \sqrt{-g^J}F(R_\mathcal{L})$, where $R_\mathcal{L} = R_J+\mathcal{L}(\phi^i,\psi^j,A_\mu^k)$ with $\mathcal{L}(\phi^i,\psi^j,A_\mu^k)$ representing the Lagrangian for all the scalars, fermions and vectors of the theory. After introducing the auxiliary field $\zeta$, this would imply having all Lagragian $\mathcal{L}(\phi^i,\psi^j,A_\mu^k)$ with a $F'(\zeta)$ prefactor (cf. eq. \eqref{eq:action:ST}). The corresponding model building and phenomenology are beyond the scope of this paper and postponed to a future work.} $R_X = R_J+X$ with $X = -g^{\mu\nu}_J \partial_\mu \phi \partial_\nu \phi$ denoting the inflaton kinetic term and $R_J = g^{\mu\nu}_J R_{\mu\nu}(\Gamma)$ where $R_{\mu\nu}(\Gamma)$ is the Ricci tensor built from the independent connection i.e. we are operating in the Palatini formulation. We can rewrite the action in the following way by introducing an auxiliary field $\zeta$:
\be
S = \int d^4x\sqrt{-g^J}\qty(\frac{F(\zeta) + F'(\zeta) (R_X -\zeta)}{2} - V(\phi)) \label{eq:action:ST} \, ,
\ee
where the symbol $'$ indicates differentiation with respect to the argument of the function. It is easy to check that action \eqref{eq:action:FRX} is obtained from action \eqref{eq:action:ST} by taking the solution of the equation of motion for $\zeta$ i.e. $\zeta = R_X$.
 By means of a conformal transformation $g_{\mu\nu}^E = F'(\zeta) g_{\mu\nu}^J$ we can rewrite the action in the Einstein frame where the theory is linear in $R$ and minimally coupled to the metric $g_{\mu\nu}^E$. The action in the Einstein frame reads:
\be \label{Einstein_action}
S = \int d^4x \sqrt{-g^E} \qty(\frac{R_E}{2} - \frac{1}{2}g^{\mu\nu}_E \partial_\mu \phi \partial_\nu \phi - U(\zeta,\phi)) \, ,
\ee
with
\be \label{eq:einstein_potential}
U(\zeta,\phi) = \frac{V(\phi)}{F'(\zeta)^2} -\frac{F(\zeta)}{2 F'(\zeta)^2} + \frac{\zeta}{2 F'(\zeta)} \, .
\ee
First of all we notice that the scalar field $\phi$ has a canonical kinetic term in the Einstein frame while $\zeta$ has no kinetic term at all, therefore it is still auxiliary also in the Einstein frame, as typical of the Palatini formulation\footnote{See \ref{appendix:A} and \ref{appendix:B} for further details.}. As we will discuss shortly, this is a crucial difference with respect to \cite{Dioguardi:2021fmr,Dioguardi:2022oqu}.
By taking the variation with respect to $\zeta$ we get the equation of motion for the auxiliary field (and the consistency condition $F'' \neq 0$):
\be \label{eq:G=V}
\frac{1}{4}\qty(2 F(\zeta) - \zeta F'(\zeta)) = V(\phi) \, ,
\ee
which in principle can be solved in $\zeta$ to get $\zeta(\phi)$. By using \eqref{eq:G=V} into \eqref{eq:einstein_potential} we get the potential in terms of the auxiliary field only 
\be\label{eq:auxilary_potential}
U(\zeta)= \frac{\zeta}{4 F'(\zeta)} \, .
\ee
Such a result has an immediate consequence on the allowed values of $\zeta$. 
When written in the form of eq. \eqref{eq:action:ST}, it is easy to understand which consistency constraints are imposed on the theory and its parameters: $F'(\zeta)>0$ is necessary to achieve a low-energy GR limit and a well defined conformal transformation. Hence, in order to have a stable positive potential suitable for inflation (see action \eqref{Einstein_action} and eq. \eqref{eq:auxilary_potential}) $\zeta \geq 0$ must hold. 

Before concluding we remark that eq. \eqref{eq:G=V} was already introduced in \cite{Dioguardi:2021fmr} where it holds as an approximation valid in the slow-roll regime\footnote{See \ref{appendix:A} for further details.}. However, in this case eq. \eqref{eq:G=V} is exact and valid even in presence of arbitrary large kinetic terms for the scalar field $\phi$. In other words, the auxiliary field $\zeta = \zeta(\phi)$ is a function of the scalar field only and not of its derivatives. This happens because the inflaton kinetic term in the Einstein frame does not have anymore any $F'$ contribution. This allows us to find an explicit form for $U(\zeta(\phi))$ in terms of the canonical scalar field whenever it is possible to solve \eqref{eq:G=V} explicitly for $\zeta = \zeta(\phi)$.

\section{Quadratic $F(R,X)$}

Now we focus on a quadratic $F$, parametrized as 
\be
F(R_X) = 2\Lambda + \omega R_X + \alpha {R_X}^2 \, , \label{eq:FRX:2}
\ee
with $\Lambda, \omega, \alpha$ real constants. This is the most general form for a quadratic $F(R_X)$. 

In such a case eq. \eqref{eq:G=V} takes the simple form

\be
\Lambda + \frac{\omega}{4}  \zeta= V(\phi) \, , \label{eq:G=V:2}
\ee
with exact solution
\be
 \zeta = \frac{4}{\omega} \left(V(\phi) -\Lambda \right) = \frac{4 \bar V(\phi)}{\omega} \, , \label{eq:zeta}
\ee
where $\bar V(\phi) =V(\phi) - \Lambda$. 
This leads to the Einstein frame potential
\be \label{eq:quadratic_einstein_2}
U(\phi) = \frac{ V(\phi) - \Lambda}{8\alpha (V(\phi)-\Lambda) +\omega^2} = \frac{\bar V(\phi)}{8\alpha \bar V(\phi) +\omega^2} \, .
\ee
The corresponding vacuum (i.e. $V(\phi)=0$) solutions are
\be \label{eq:vacuum}
\zeta_0= -4\Lambda/\omega , \qquad  U_\Lambda = \frac{\Lambda}{8\alpha \Lambda -\omega^2} \, .
\ee
Moreover, the large potential configuration (
$V(\phi)\to\pm\infty$) presents a plateau at (see Figs. \ref{fig.plot.U.pos}, \ref{fig.plot.U.neg})

\be
U \approx U_\alpha=\frac{1}{8 \alpha} \, . \label{eq:U:alpha} 
\ee

Requiring the positivity of $\zeta$ and $U(\phi)$ at any $\phi$ values, vacuum configuration included (i.e. $\zeta_0, U_\Lambda \geq 0$) leads to two possible scenarios:
\bea
&& 1) \quad \omega >0, \quad \Lambda \leq 0, \quad \bar V \geq 0 \, , 
\label{eq:w>0}\\
&& 2) \quad \omega <0, \quad \Lambda >0, \quad \bar V < 0 \, , 
\label{eq:w<0}
\eea
with the common constraint 
\be
\alpha > \frac{\omega^2}{8 \Lambda} \, \label{eq:alpha:lowerbound}
\ee
only in case $\Lambda \neq 0$.
Notice that an exactly null $\Lambda$ (or alternatively $\zeta_0=0$) is allowed only for $\omega>0$. 
The sign of $\omega$ affects the behaviour of the Einstein frame potential \eqref{eq:quadratic_einstein_2} in a way that will be clarified later on.
In the following we will study the two different sign choices separately.

\subsection{The standard case: $\omega >0$}

First of all we notice that, if $V(\phi)$ has an absolute minimum (or alternatively a horizontal asymptote), we can set it to be null via a redefinition of $\Lambda$ as long as \eqref{eq:w>0} is satisfied. Without loss of generality, in this subsection we work in this configuration. Morever, eq. \eqref{eq:w>0} would in principle allow for negative values for $\alpha$. However, unless some \emph{ad hoc} choice of $V(\phi)$, negative values for $\alpha$ would also imply the insurgence of a pole in $U(\phi)$. In order to avoid such a case, we restrict ourselves to $\alpha>0$. In such a case, given the constraints \eqref{eq:w>0}, we can easily check that $U_\alpha > U_\Lambda$ and $U_\Lambda$ becomes the cosmological constant $U_\text{CC}$. Consistency with data \cite{Planck:2018vyg} then requires $U_\Lambda = U_\text{CC} \sim 10^{-47} \text{GeV}^{4}$. Treating $\alpha$ and $\omega$ as free parameters, such a constraint can be solved in function of $\Lambda$, giving 
\begin{equation}
    \Lambda = \frac{\omega^2 \, U_{\text{CC}}}{8 \alpha  U_{\text{CC}}-1} . \label{eq:Lambda:CC}
\end{equation}
Unless we choose ad hoc gigantic values for $\alpha$ and/or $\omega$, it is easy to check that $|\Lambda|$ must be approximately of the same order of magnitude as $U_\text{CC}$.
Therefore this setup cannot actually solve the cosmological constant problem, but only reproduce its value by tuning the free parameters of the model.
\begin{figure}[t]
    \centering
    \includegraphics[width=0.45\textwidth]{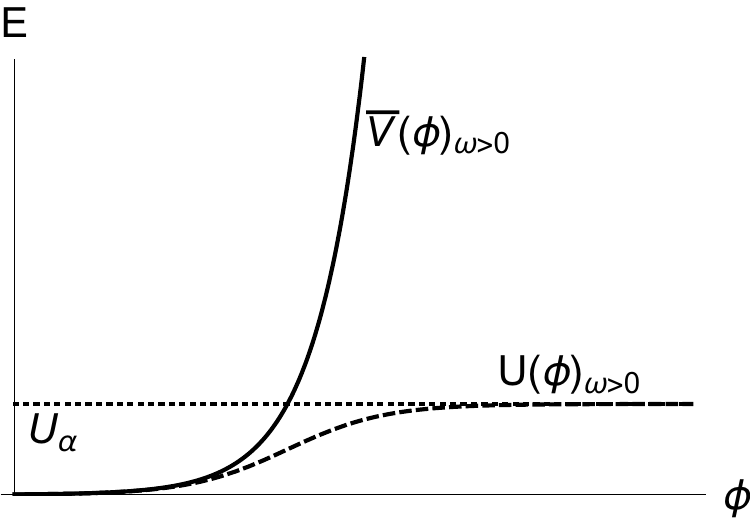}
    \caption{Reference plots for $\bar V(\phi)$ (continuous) and $U(\phi)$ (dashed) vs. $\phi$ when $\omega>0$.}
    \label{fig.plot.U.pos}
\end{figure}
Comparing our result \eqref{eq:quadratic_einstein_2} with the one in \cite{Enckell:2018hmo}, we notice that we retain the same form of the potential (i.e. a plateau for big enough $\bar V(\phi)$, see Fig. \ref{fig.plot.U.pos}). In particular, at large potential configuration $V(\phi) \rightarrow \infty$, this plateau takes the form:
\be \label{eq:Einstein_potential_expanded}
U(\phi) \approx U_\alpha \qty(1-\frac{\omega^2 U_\alpha}{V(\phi)})
\ee
Hence, we can provide an asymptotically flat potential independently of the original choice of $V(\phi)$ as in the standard $F(R)$ case \cite{Enckell:2018hmo} but with the advantage of having a canonically normalized scalar field $\phi$ and no higher order kinetic term.  

In order to describe the inflationary predictions, we need first of all the slow roll-parameters, defined as:
\bea
\epsilon  (\phi) &=& \frac{1}{2}\qty(\frac{U'(\phi)}{U(\phi)})^2 \, , \label{eq:epsilon}
\\
\eta  (\phi) &=& \frac{U''(\phi)}{U(\phi)} \, . 
\eea
The first slow-roll parameter $\epsilon$ allows us to compute the number of e-folds of expansion of the Universe as
\be
N_e =  \int_{\phi_{\textrm{end}}}^{\phi_N} {\rm d}\phi \, \frac{U(\phi)}{U'(\phi)} ,
\label{eq:Ne}
\ee
where the field value at the end of inflation\footnote{One might argue that also $|\eta|=1$ could trigger the end of slow-roll before reaching $\epsilon = 1$. However, this is never the case in our example scenarios, at least for the parameter space that we considered.} is given by $\epsilon  (\phi_{\textrm{end}}) = 1 $, while the field value $\phi_N$ at the time a given scale left the horizon is given by the corresponding $N_e$. The spectral index $n_\textrm{s}$ and the tensor-to-scalar ratio $r$ are:
\bea
n_\textrm{s}  &=& 1+2\eta  (\phi_N)-6\epsilon  (\phi_N) \, ,  \label{eq:ns} \\
r  &=& 16\epsilon  (\phi_N) \,  , \label{eq:r}
\eea
and the amplitude of the scalar power spectrum \cite{Planck2018:inflation} is 
\be
 A _\textrm{s} = \frac{1}{24 \pi^2}\frac{U(\phi_N)}{\epsilon  (\phi_N)} \simeq 2.1 \times 10^{-9} \, .
 \label{eq:As:th}
\ee
\begin{figure*}[t]
    \centering
    
    \subfloat[]{\includegraphics[width=0.45\textwidth]{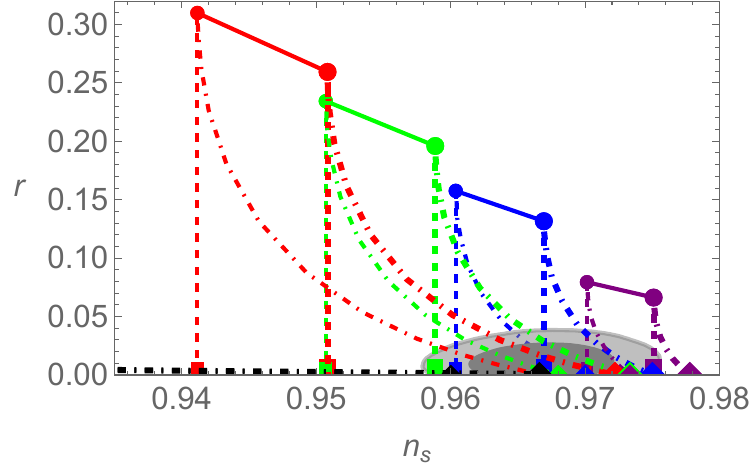}}%
    \qquad
    \subfloat[]{\includegraphics[width=0.45\textwidth]{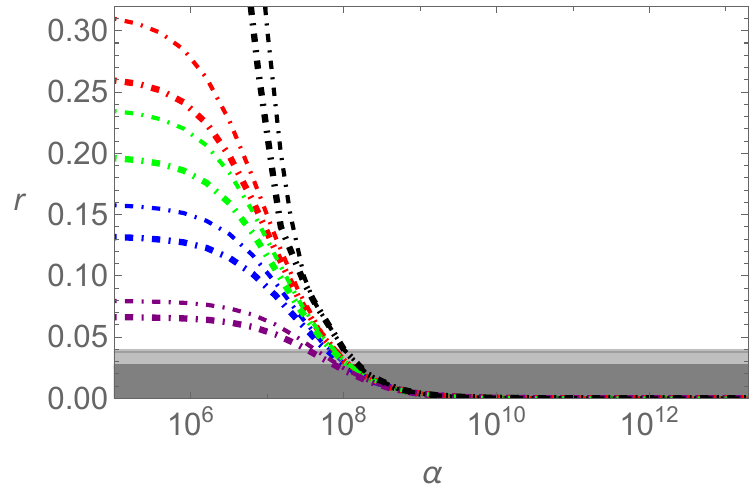}}%
    
    \subfloat[]{\includegraphics[width=0.45\textwidth]{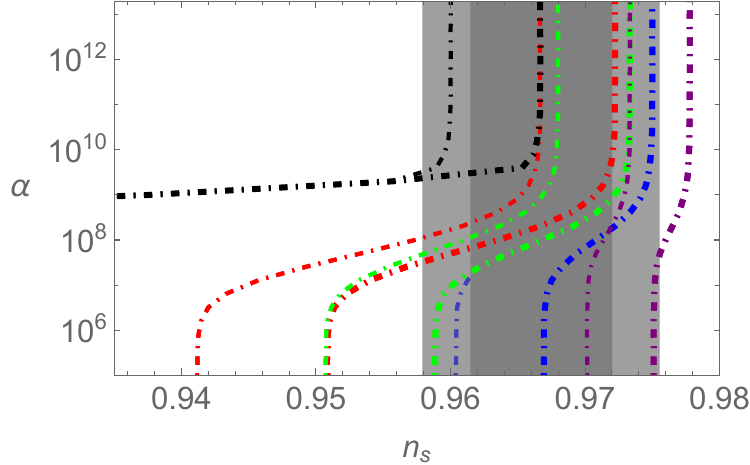}}%
    \qquad
    \subfloat[]{\includegraphics[width=0.45\textwidth]{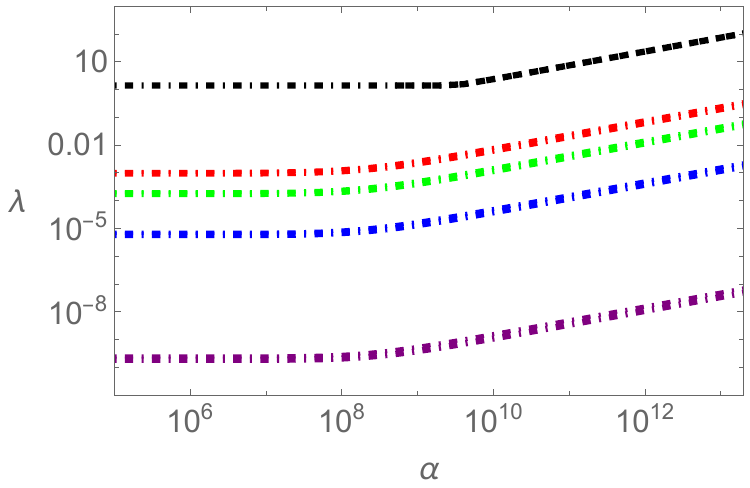}}
    \caption{$r$ vs. $n_s$ (a), $r$ vs. $\alpha$ (b), $\alpha$ vs. $n_s$ (c), $\lambda$ vs. $\alpha$ (d) for $\omega=1$ and $V(\phi) = \lambda_k \phi^k$ with $k=1$ (purple), $k=2$ (blue), $k=3$ (green), $k=4$ (red) and $k \to \infty$ (black) for $N_e = 50$ (thin, dot-dashed line) and $N_e = 60$ (thick, dot-dashed line). In the same color code we show the limiting values $\alpha=0$ (continuous line, bullets) and $\alpha \rightarrow \infty$ (triangles). In (a) the dashed lines represent the prediction of the standard Palatini $R^2$ model \cite{Enckell:2018hmo} for the corresponding potential and the squares the limiting values for $\alpha\rightarrow \infty$.  The gray areas represent the 1,2$\sigma$ allowed regions coming  from  the latest combination of Planck, BICEP/Keck and BAO data \cite{BICEP:2021xfz}. }
    \label{r_vs_ns_R2.png}
\end{figure*}
In the standard quadratic $F(R)$ case \cite{Enckell:2018hmo}, at the leading order, only $r$ is affected by the Palatini $R^2$ term, while all the other observables stay unchanged. In contrast to that, here we cannot find a compact straightforward expression for the phenomenological parameters in eqs. \eqref{eq:ns}, \eqref{eq:r} and \eqref{eq:As:th} in terms of the corresponding ones computed for $V(\phi)$, because now the inflaton field $\phi$ is already canonically normalized in the Einstein frame action \eqref{Einstein_action}.
However we can still find a compact expression for $r$ in the strong coupling limit $\alpha \to \infty$. In such a case, the Einstein frame potential asymptotically approaches the plateau  \eqref{eq:U:alpha}. Therefore, inserting this value in \eqref{eq:As:th} we can easily prove that at the leading order
\be
 r  \approx \frac{1}{12 \pi^2 A_s \alpha} \, . \label{eq:r:limit}
\ee
This means that we can arbitrarily lower $r$ by increasing $\alpha$, exactly as found in \cite{Enckell:2018hmo}. On the contrary, we cannot make a model independent prediction for $n_s$.

For illustrative purposes, we consider a monomial potential of the form 
\be
V(\phi) = \lambda_k \phi^k, \qquad \lambda_k = \frac{\lambda^k}{k!} \label{eq:V:k}
\ee
where the unusual prefactor, $\lambda_k$, is chosen for numerical convenience.
In such a case, we have:
\bea
N_e &=& \int^{\phi_N}_{\phi_{end}} d\phi \frac{1}{\sqrt{2} \sqrt\epsilon_+ (\phi) }
\\
r  &=& 16 \epsilon_+ (\phi_N) 
\\
n_s &=& 1 - 6\epsilon_+ (\phi_N) + 2\eta_+ (\phi_N)
\\
A_s &=& \frac{\omega^2 \Bar V(\phi_N)}{24\pi^2 \epsilon_+(\phi_N) \left(8\alpha\Bar V(\phi_N)+\omega^2\right)}
\eea
where 
\bea
 &&\hspace*{-1cm} \epsilon_+(\phi) = \frac{\lambda_k^2 k^2 \omega^4 \phi^{2k-2}}{2\Bar V(\phi)^2(8\alpha\Bar V(\phi) +\omega^2)^2} \\
&&\hspace*{-1cm} \eta_+(\phi) = \frac{2 \lambda_k \omega^2 \phi^{k-2}}{\Bar V(\phi)(8\alpha\Bar V(\phi)+\omega^2)^2} \times\\  && \qty(8\alpha(k+1)\lambda_k\phi^k + (k-1)(8\alpha\Lambda-\omega^2))
\eea
 In the strong coupling limit $\alpha \rightarrow \infty$, we recover the well-known polynomial $\alpha$-attractors \cite{Kallosh:2022feu}, whose predictions are:
\bea
r  &\sim& 0
\\
n_s  &=& 1-\frac{k+1}{k+2}\frac{2}{N_e} \label{eq:ns:plus}
\eea
A numerical analysis for the reference value $\omega = 1$ with $\Lambda$ and $\lambda$ fixed using $\lambda$ eqs. \eqref{eq:Lambda:CC} and \eqref{eq:As:th}  is showed in Fig. \ref{r_vs_ns_R2.png}, where we plot the observables $r$ vs. $n_s$ (a), $r$ vs. $\alpha$ (b), $\alpha$ vs. $n_s$ (c), $\lambda$ vs. $\alpha$ (d) for $V(\phi) = \lambda_k \phi^k$ with $k=1$ (purple), $k=2$ (blue), $k=3$ (green), $k=4$ (red) and $k \to \infty$ (equivalent to $V(\phi)=e^{\lambda \phi}$) (black) for $N_e = 50$ (thin, dot-dashed line) and $N_e = 60$ (thick, dot-dashed line). In the same color code we show the limiting values\footnote{The case $k \to \infty$ for $\alpha=0$ is not visible in Fig. \ref{r_vs_ns_R2.png}(a) because far away from the allowed region.} $\alpha=0$ (continuous line, bullets) and $\alpha \rightarrow \infty$ (triangles). In (a) the dashed lines represent the prediction of the standard Palatini $R^2$ model \cite{Enckell:2018hmo} for the corresponding potential and the squares the limiting values for $\alpha\rightarrow \infty$. The gray areas represent the 1,2$\sigma$ allowed regions coming  from  the latest combination of Planck, BICEP/Keck and BAO data \cite{BICEP:2021xfz}. 
Comparing the results of the quadratic $R_X$ model with the standard one \cite{Enckell:2018hmo}, we notice that, as expected, we still have a suppression in $r$. On the other hand, while in  \cite{Enckell:2018hmo} $n_s$ stays essentially unchanged, now it increases by increasing $\alpha$. In this way, for any $k \geq 1$, it is possible to find a $N_e$ value so that the predicted $r$ and $n_s$ lay in the experimentally allowed region \cite{BICEP:2021xfz}. Finally, in (d) we notice that the coupling $\lambda$ has to increase along with $\alpha$ in order to satisfy the constraint \eqref{eq:As:th}. Notice in particular that the value of $\lambda$ does not change substantially between $N_e = 50$ and $N_e =60$ so the two lines appear almost superimposed in the plot.

To conclude we stress that the form of \eqref{eq:Einstein_potential_expanded} is the same as in \cite{Kallosh:2022feu} but our construction is certainly more immediate as it does not require any field redefinition,  and more general because it can be applied to any positive $V$, not just monomial choices. Since the potential in \eqref{eq:quadratic_einstein_2} with the setup \eqref{eq:w>0} predicts asymptotically flat potentials whenever $\alpha \gg \omega$, we label such a choice as \emph{canonical fractional attractors}, to be counterposed to a more \emph{exotic} configuration which will be discussed in the next subsection.

\subsection{The alternative case: $\omega<0$}   

\begin{figure}[t]
    \centering
    \includegraphics[width=0.45\textwidth]{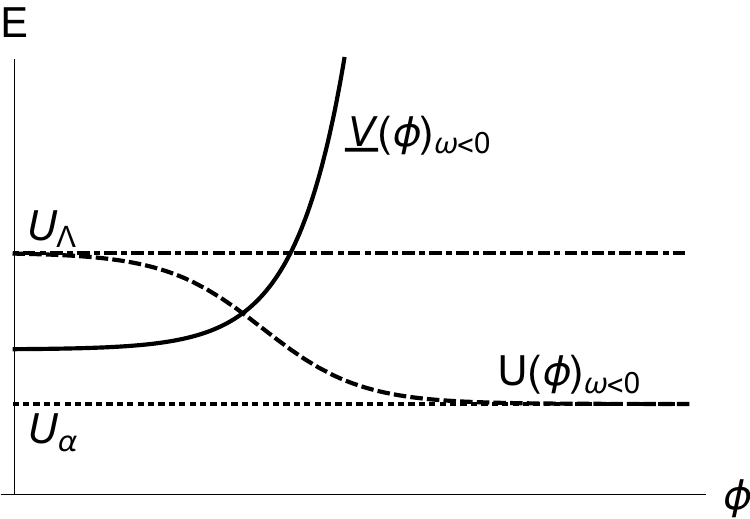}
    \caption{Reference plots for $\underline V(\phi)$ (continuous) and $U(\phi)$ (dashed) vs. $\phi$ when $\omega<0$.}
    \label{fig.plot.U.neg}
\end{figure}

The first peculiar feature of this configuration is the $\emph{wrong}$ sign for the linear term in $R$ in the Jordan frame action \eqref{eq:action:ST}. At first glance it might seem that such a configuration is forbidden, coming with issues in the Jordan frame like repulsive gravity or a negative inflaton kinetic energy. However, as explained after eq. \eqref{eq:auxilary_potential}, the condition $F'>0$ ensures that gravity stays attractive. The same holds for the inflaton kinetic term, which receives not only contributions from the linear term in $\omega$ but also from the quadratic term $\alpha R_X^2$ (see eq. \eqref{eq:FRX:2}). When all the contributions are taken into account, it is easy to prove that the positivity of $F'$ ensures also the positivity of the inflaton kinetic energy (see also eq. \eqref{eq:action:phizeta:J}). Therefore such a setup is allowed as long as the consistency constraint of $F'>0$  is satisfied (provided that $F''\neq 0$) (e.g. \cite{Dioguardi:2021fmr,Dioguardi:2022oqu} and refs. therein). This happens when all the conditions in eq. \eqref{eq:w<0} are respected. Particularly relevant is the role of $\Lambda>0$ and of the lower bound on $\alpha$ in eq. \eqref{eq:alpha:lowerbound} that ensure the positivity of $F'$ wherever eq. \eqref{eq:zeta} admits a solution. The simplest way\footnote{Another possible, but more tuned, choice is to have $V(\phi)$ positive but bounded from above in such a way that $\bar V$ remains negative.} to achieve it is to have the Jordan frame scalar potential $V(\phi)$ (semi)definite negative. Analogously to the previous case, also now $\alpha$ needs to be constrained to positive values. It is convenient to introduce 
\be
\underline V(\phi) = - \bar V(\phi)= - V(\phi) + \Lambda \ > \frac{\omega^2}{8 \alpha}\, , \label{eq:underline:V}
\ee
which is strictly positive  and bounded from below because of \eqref{eq:w<0}. In this way we can rewrite the Einstein frame potential as
\be \label{eq:quadratic_einstein_tail}
U(\phi)  = \frac{\underline V(\phi)}{8\alpha \underline V(\phi) -\omega^2} \, .
\ee
Comparing with eq. \eqref{eq:quadratic_einstein_2}, now $\omega^2$ contributes with a negative sign to the denominator. However it can be easily proven that the Einstein frame potential \eqref{eq:quadratic_einstein_tail} is positive and bounded from above (without any appearance of a pole) thanks to \eqref{eq:w<0} and \eqref{eq:underline:V}. A reference plot for $U(\phi)$ is given in Fig. \ref{fig.plot.U.neg}. We notice that now the potential exhibits two plateaus, the usual $U_\alpha$ in \eqref{eq:U:alpha}, when $\underline V \to \infty$, but also another one, $U \approx U_\Lambda$ in \eqref{eq:vacuum}, when $\underline V \to \Lambda$ i.e. $V \to 0$. Notice that in opposition to the previous $\omega>0$ case, given \eqref{eq:w<0} and \eqref{eq:underline:V}, now $U_\Lambda > U_\alpha$. The height of the plateaus is dictated only by the $F$'s parameters $\Lambda, \omega, \alpha$ regardless\footnote{If we would consider the Einstein frame action \eqref{Einstein_action} with $U(\phi)$ given by \eqref{eq:quadratic_einstein_tail} as a starting point, then all the dependence on the original parameter $\Lambda$ would be moved to the existence of a strictly positive minimum value for $\underline V$.}
of the initial choice of $V(\phi)$. We label such class of potentials as \emph{tailed-fractional attractors}. Since we have two plateaus, it is in principle possible to use one single potential to explain both early universe inflation, which happens close to the $U_\Lambda$ plateau, and late universe acceleration, where the value $U_\alpha$ sets the value of the cosmological constant observed today. We will discuss such a topic later on. Now we focus on the phenomenology around the first plateau. In its proximity (in the strong coupling regime), the scalar potential can be approximated as:
\be
U(\phi) = U_\Lambda \qty(1-\frac{U_\Lambda}{\Lambda U_\alpha}\abs{V(\phi)}) \, ,
\ee
which has the form of a generalized hilltop potential.
Also in this case we can get a universal limit for the tensor-to-scalar ratio $r$ only
\be
 r \approx \frac{2}{3 \pi^2 A_s} \frac{\Lambda}{8\Lambda \alpha - \omega^2} \, , \label{eq:r:limit:tail}
\ee
which holds around the first plateau, that is for $V \sim 0$.
When $\alpha \gg 1$, the most trivial configuration is the one where $\omega$ can be neglected and eq. \eqref{eq:r:limit:tail} reduces to eq. \eqref{eq:r:limit}. 

\begin{figure}[t]
    \includegraphics[width=0.45\textwidth]{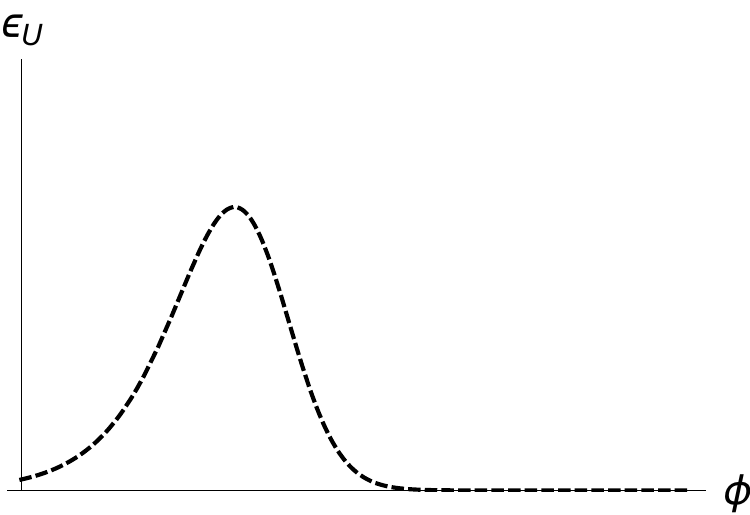}
    \caption{Reference plot for $\epsilon  (\phi)$ (dashed) vs. $\phi$ when $\omega<0$.}
    \label{fig.plot.eps.neg}
\end{figure}

Before moving to a more quantitative analysis, it is important to stress that an additional constraint needs to be imposed in order to achieve the end of slow-roll. A qualitative plot for the first slow-roll parameter $\epsilon  $ is shown in Fig. \ref{fig.plot.eps.neg}. As expected, for $V(\phi) \to 0$  (or $-\infty$), $\epsilon   \to 0$ since $U(\phi)$ approaches the corresponding plateaus. However, in between the two plateaus, $\epsilon  $ does not diverge to $+\infty$, ensuring a certain end of slow-roll, but it reaches a local maximum. This is because the absolute minimum of $U(\phi)$ is not null. Therefore, in order to end slow-roll, we need to ensure that the local maximum in $\epsilon  $ is bigger than one. We can intuitively understand (cf. eq. \eqref{eq:epsilon}) that this is achieved by lowering the second plateau $U \approx \frac{1}{8\alpha}$ i.e. by increasing $\alpha$. Therefore this sets a minimum value $\alpha>\alpha_{min}$. Its numerical value is model dependent and must be calculated for the specific choice of the scalar potential $V(\phi)$. However, the existence of the minimum value $\alpha_{min}$ is model independent. 

   \begin{figure*}[t]
    \centering
    
    \subfloat[]{\includegraphics[width=0.45\textwidth]{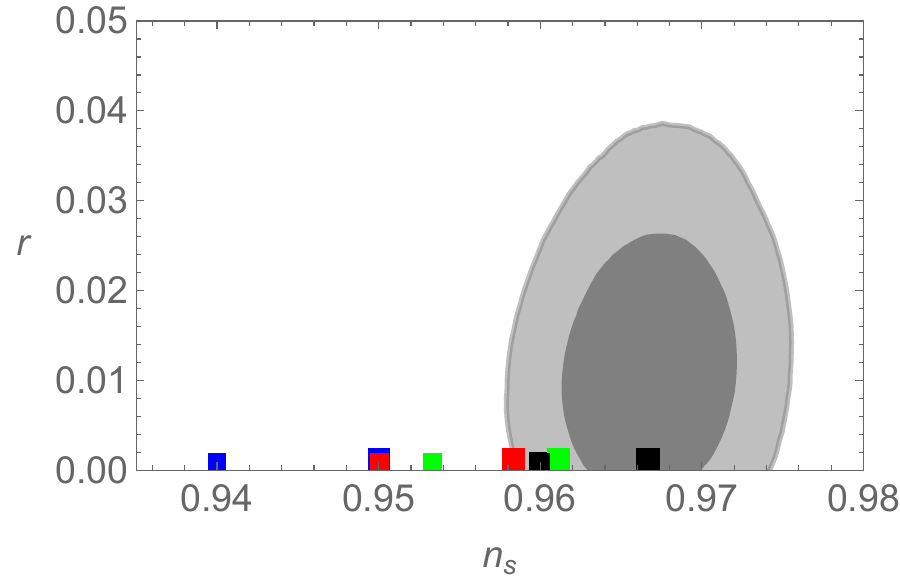}}%
    \qquad
    \subfloat[]{\includegraphics[width=0.45\textwidth]{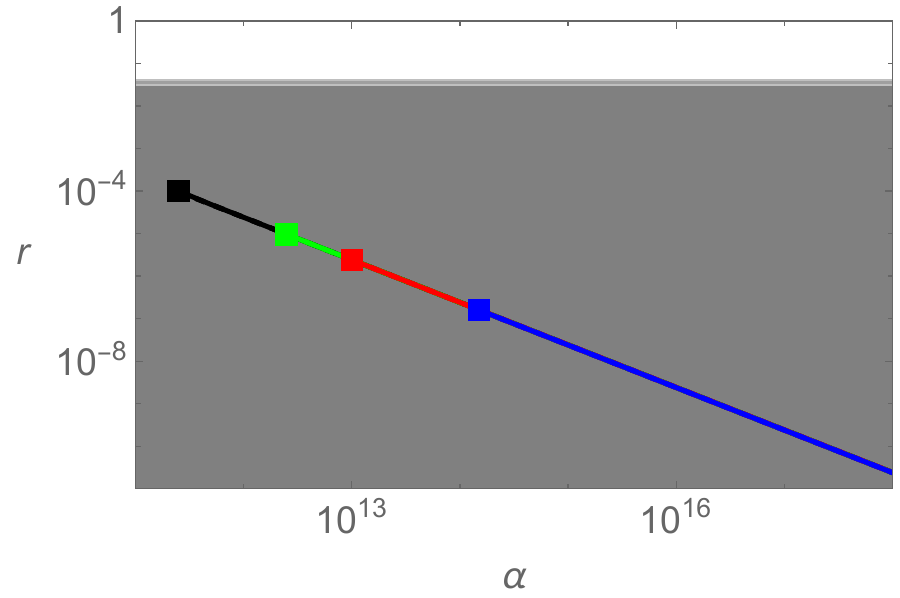}}%
    
    \subfloat[]{\includegraphics[width=0.45\textwidth]{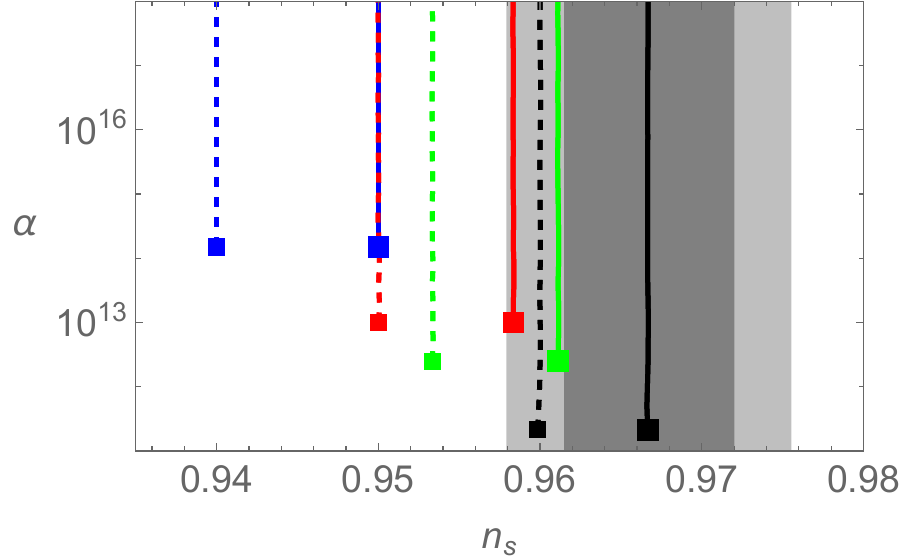}}%
    \qquad
    \subfloat[]{\includegraphics[width=0.45\textwidth]{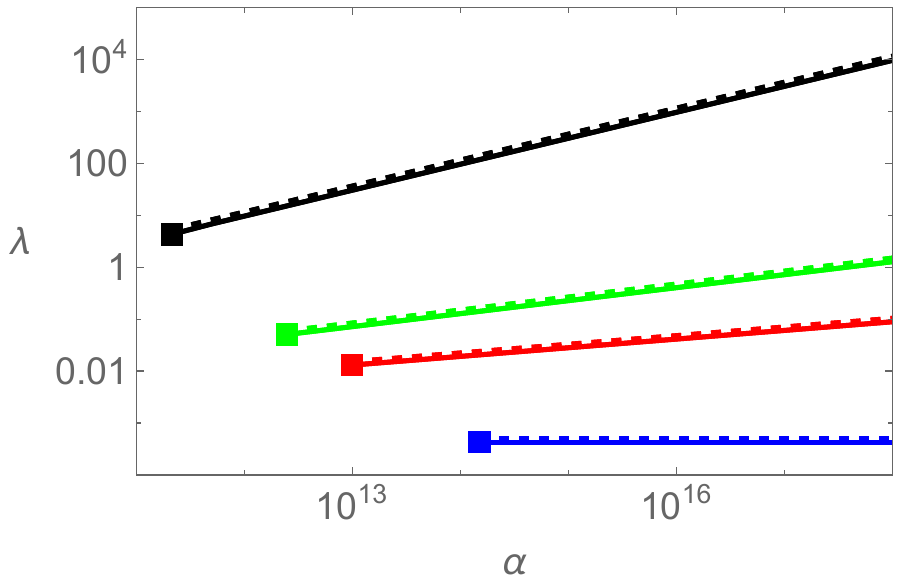}}
    \caption{$r$ vs. $n_s$ (a), $r$ vs. $\alpha$ (b), $\alpha$ vs. $n_s$ (c), $\lambda$ vs. $\alpha$ (d) for $\omega=-1$ and $V(\phi) = -\lambda_k \phi^k$  with $k=4$ (blue), $k=6$ (red), $k=8$ (green) and $k \to \infty$ (black) for $N_e = 50$ (dashed), and $N_e = 60$ (continuous). In the same color code the squares represent the value $\alpha_{min}$ for which we can have a graceful exit from inflation. The grey areas represent the 1,2$\sigma$ allowed regions coming  from  the latest combination of Planck, BICEP/Keck and BAO data \cite{BICEP:2021xfz}.}
    \label{r_vs_ns_tail_R2.png}
\end{figure*}
In order to perform a more quantitative comparison with the canonical fractional attractors case, we consider now a negative monomial potential $V(\phi) = - \lambda_k \phi^k$ with $k \geq 3$ and $\lambda_k$ defined in eq. \eqref{eq:V:k}. In this case the inflationary observables are:

\bea
N_e &=& \int_{\phi_N}^{\phi_{end}} d\phi \frac{1}{\sqrt{2} \sqrt\epsilon_-(\phi) }
\\
r  &=& 16 \epsilon_- (\phi_N)
\\
n_s &=& 1 - 6\epsilon_- (\phi_N) + 2\eta_- (\phi_N)
\\
A_s &=& \frac{\omega^2 \underline V(\phi_N)}{24\pi^2 \epsilon_-(\phi_N) \left(8\alpha\underline V(\phi_N)-\omega^2 \right)}
\eea
where
\bea
 &&\hspace*{-1cm} \epsilon_- (\phi) = \frac{\lambda_k^2 k^2 \omega^4 \phi^{2k-2}}{2\underline V(\phi)^2(8\alpha\underline V(\phi)-\omega^2)^2} \\
&&\hspace*{-1cm}\eta_- (\phi) = \frac{2 \lambda_k \omega^2 \phi^{k-2}}{\underline V(\phi)( 8\alpha\underline V(\phi)-\omega^2)^2} \times\\
&& \qty(8\alpha(k+1)\lambda_k\phi^k + (k-1)(8\alpha\Lambda-\omega^2))
\eea
The corresponding strong coupling results are:
\bea
r &\sim &0 \\
n_s &=& 1 - \frac{k-1}{k-2}\frac{2}{N_e} \, , \label{eq:ns:minus}
\eea
which are the hilltop inflationary predictions for small $r$, as expected. 

 A numerical study for the reference value $\omega = -1$ with $\Lambda \simeq \frac{6}{5} \frac{1}{8 \alpha}$, ensuring that eq. \eqref{eq:alpha:lowerbound} holds, and $\lambda$ fixed using eq. \eqref{eq:As:th} is showed in Fig. \ref{r_vs_ns_tail_R2.png}, where we plot the observables $r$ vs. $n_s$ (a), $r$ vs. $\alpha$ (b), $\alpha$ vs. $n_s$ (c), $\lambda$ vs. $\alpha$ (d) for $V(\phi) = -\lambda_k \phi^k$  with $k=4$ (blue), $k=6$ (red), $k=8$ (green) and $k \to \infty$ (black) for $N_e = 50$ (dashed), and $N_e = 60$ (continuous). In the same color code the squares represent the value $\alpha_{min}$ for which we can have a graceful exit from inflation. The grey areas represent the 1,2$\sigma$ allowed regions coming  from  the latest combination of Planck, BICEP/Keck and BAO data \cite{BICEP:2021xfz}. In this case we see that agreement with data requires $k \geq 6$.
As in the previous case, from (d) we notice that the coupling $\lambda$ has to increase along with $\alpha$ in order to satisfy the constraint \eqref{eq:As:th}. Notice in particular that the value of $\lambda$ does not change substantially between $N_e = 50$ and $N_e =60$ so the two lines appear almost superimposed in the plot.
Moreover, we notice that for $k \to \infty$ (equivalent to replacing $\lambda_k \phi^k$ with $e^{\lambda \phi}$), \eqref{eq:ns:plus} and \eqref{eq:ns:minus} converge to the same limit.
Before concluding, we stress that the model predicts $r\lesssim 10^{-4}$, implying that the predicted tensor-to-scalar ratio would not be measurable even by more futuristic high-resolution satellite mission, such as PICO \cite{NASAPICO:2019thw}. This happens because the requirement $\alpha > \alpha_{min}$, needed to achieve the end of slow-roll, automatically forces $\alpha$ in the strong coupling regime.

 Moreover, when $\alpha > \alpha_{min}$, not only slow-roll inflation ends but also restarts at a later time, allowing the possibility for a joined explanation of the earlier and later accelerated expansion of the Universe, as already mentioned before. In order to understand the corresponding phenomenology, it is convenient to introduce the parameter $\delta$ so that
 
 \be\label{eq:delta}
 \alpha\Lambda = \frac{\omega^2}{8}(1+\delta).
 \ee
 Using such a definition in \eqref{eq:r:limit:tail}, we obtain
\be
 r \approx \frac{1}{12 \pi^2 A_s \alpha} \frac{1+\delta}{\delta} \, . \label{eq:r:limit:tail:2}
\ee
in the limit of $\alpha \gg 1$.
Also, as long as $\delta \sim O(1)$ (in Fig. \ref{r_vs_ns_tail_R2.png} we used $\delta \simeq 0.2$), the exact numerical value for $r$ depends on the actual value of $\delta$ (unless $\delta \ll 1$), but it is still suppressed and of the same order of \eqref{eq:r:limit} in the big $\alpha$ limit. In this case, however, it is immediate to see that:
\be
U_\Lambda = \frac{\Lambda}{8\alpha\Lambda-\omega^2} = \frac{1}{8\alpha}\frac{1+\delta}{\delta} \gtrsim \frac{1}{8\alpha} \, .
\ee
Therefore, adjusting $U_\alpha=\frac{1}{8\alpha}$ to the observed value of the vacuum energy density \cite{Planck:2018vyg} also lowers the inflationary plateau making its value too low to be phenomenologically consistent with the evolution of the universe.  An alternative option is to take $\delta \ll 1$, but $\alpha \delta A_s \gg 1$ so that we still get a small value
\be
r \sim \frac{1}{12\pi^2 \alpha \delta A_s } \, .
\ee

For instance, if we consider $U_\alpha = \frac{1}{8\alpha} =  U_\text{CC} \sim 10^{-47} \text{GeV}^{4}$
we have that $\alpha \sim 10^{122}$. In order to get a value of $r\sim 10^{-6}$ (which still corresponds to a high enough energy scale for inflation around $10^{15} \text{GeV}$) we would need an extremely fine-tuned $\delta \sim 10^{-110}$. 
Unfortunately, even though mathematically possible, such a tuning of the parameters, looks more like a confirmation rather than a solution of the cosmological constant problem.

On the other hand, it has been proven that quadratic $F$'s with negative $\omega$, are the strong coupling limit configuration for $F$'s of order higher than quadratic \cite{Dioguardi:2022oqu}. In such cases, independently of the chosen $F$, the energy potential tail automatically approaches zero by construction, resembling the feature of the quintessential inflation models (e.g. \cite{de_Haro_2021} and refs. therein), implying a possible more natural solution of the cosmological constant problem. However, the search for a successful quintessential $F(R_X)$ model is beyond the scope of the present article and will be postponed to a future work.

\section{Conclusions}

We studied single-field inflation embedded in $F(R_X)$ Palatini gravity. We showed that such a choice solves several of the issues araising from the usual Palatini $F(R)$ models while still providing flat inflaton potentials. In particular we focused on quadratic $F(R_X)$'s showing that only two possible configurations are possible. Those configurations both lead to flat inflaton potentials that can be classified as attractors given their general prediction on the inflationary observables. We named those two classes canonical fractional attractors, which generalize the well known polynomial $\alpha$-attractors, and tailed fractional-attractors which generalize the predictions of hilltop potentials. Moreover, we showed that the class of quadratic theories with $\omega<0$ can in principle explain both early and late universe accelerated expansion with one single potential.  Even though fine tuned, this is a quite intriguing finding. In light of the results of \cite{Dioguardi:2022oqu}, such an idea deserves further studies, going to appear in a future work.

\section*{Acknowledgments}

This work was supported by the Estonian Research Council grants PRG1055,  RVTT3, RVTT7 and the CoE program TK202 ``Fundamental Universe'’.

\appendix

\section{Palatini $F(R)$ and its issues} \label{appendix:A}
In this appendix we give a brief summary of the results and issues found in the context of slow-roll inflation embedded in Palatini $F(R)$ gravity. More details can be found in \cite{Dioguardi:2021fmr,Dioguardi:2022oqu}. 
We start by taking the action 
\be
S = \int d^4x \sqrt{-g^J}\qty(\frac{1}{2}F(R_J) - \frac{1}{2} g^{\mu\nu}_J \partial_\mu \phi \partial_\nu \phi - V(\phi)) \label{eq:action:FR}
\ee
where we assumed Planck units, $\MP=1$, a space-like metric signature. $V(\phi)$ is the inflaton scalar potential and $F(R)$ is an arbitrary function of its argument with $R_J = g^{\mu\nu}_J R_{\mu\nu}(\Gamma)$ where $R_{\mu\nu}(\Gamma)$ is the Ricci tensor built from the independent connection i.e. we are operating in the Palatini formulation. We can rewrite the action by introducing an auxiliary field $\zeta$:
\bea
&&\hspace{-1.4cm} S = \int d^4x\sqrt{-g^J}\Bigg(\frac{F(\zeta) + F'(\zeta) (R_J -\zeta)}{2} \nn\\
&&- \frac{1}{2} g^{\mu\nu}_J \partial_\mu \phi \partial_\nu \phi - V(\phi) \Bigg) \label{eq:action:ST:FR} \, ,
\eea
where the symbol $'$ indicates differentiation with respect to the argument of the function. It is easy to check that the action \eqref{eq:action:FR} is obtained from the action \eqref{eq:action:ST:FR} by using the solution of the equation of motion for $\zeta$ i.e. $\zeta = R_J$.
 By means of a conformal transformation $g_{\mu\nu}^E = F'(\zeta) g_{\mu\nu}^J$ we can rewrite the action in the Einstein frame where the theory is linear in $R$ and minimally coupled to the metric $g_{\mu\nu}^E$. The action in the Einstein frame reads:
\be \label{Einstein_action:FR}
S = \int d^4x \sqrt{-g^E} \qty(\frac{R_E}{2} - \frac{g^{\mu\nu}_E \partial_\mu \phi \partial_\nu \phi}{2  F'(\zeta)}  - U(\zeta,\phi)) \, ,
\ee
with
\be \label{eq:einstein_potential:FR}
U(\zeta,\phi) = \frac{V(\phi)}{F'(\zeta)^2} -\frac{F(\zeta)}{2 F'(\zeta)^2} + \frac{\zeta}{2 F'(\zeta)} \, .
\ee
Note that $U(\zeta,\phi)$ as the same functional form as for the $F(R_X)$ case shown in eq. \eqref{eq:einstein_potential}.
Provided the consistency condition $F'' \neq 0$, the equations of motion of the system are (in the approximation of exactly homogeneous and isotropic Universe):
\begin{gather}
    \label{eq:phi_eom}
    \ddot{\phi} + 3H\dot{\phi} + \frac{V'(\phi)}{F'(\zeta)} = \frac{\dot{\phi}\dot{\zeta}F''(\zeta)}{F'(\zeta)}  \, , \\
    \label{eq:H_eom}
    3H^2 = \frac{1}{2}\frac{\dot{\phi}^2}{F'(\zeta)} + U(\phi,\zeta) \, , \\
    \label{eq:zeta_eom}
    -\frac{1}{2}\dot{\phi}^2 F'(\zeta) + 2V(\phi) -  G(\zeta) = 0 \, ,
\end{gather}
with
\be
\label{eq:G:FR}
 G(\zeta) = 
\frac{1}{4}\qty(2 F(\zeta) - \zeta F'(\zeta)) \, .
\ee
It can be easily checked that eq. \eqref{eq:zeta_eom}, when plugged back into action \eqref{Einstein_action:FR} induces non trivial higher order inflaton kinetic terms which complicate a lot the dynamics of the system.
Moreover, solving exactly the equation \eqref{eq:zeta_eom} proves to be hard for a generic $F(R)$, therefore we replace it with an equation for the time derivative of $\zeta$ \cite{Dioguardi:2021fmr},
\begin{equation} \label{eq:zeta_deriv}
    \dot{\zeta} = \frac{3H\dot{\phi}^2 F'(\zeta)  + 3 V'(\phi)\dot{\phi}}{2G'(\zeta) + \frac{3}{2}\dot{\phi}^2 F''(\zeta)} \, .
\end{equation}
As long as $\zeta$ solves \eqref{eq:zeta_eom} initially, equation~\eqref{eq:zeta_deriv} guarantees that \eqref{eq:zeta_eom} holds at all times. Eq. \eqref{eq:zeta_deriv} also shows an issue that might arise in generic $F(R)$ models: if $G'(\zeta)$ and $F''(\zeta)$ have opposite signs, then $\dot\zeta$ presents a pole. Such a pole is a direct consequence of the $1/F'$ prefactor in front of the inflaton kinetic term in the Einstein frame (see eq. \eqref{Einstein_action:FR}) and it might bring to catastrophic consequences for the dynamics of the system \cite{Dioguardi:2022oqu}. For instance, this happens when $F$ is a function of higher order than quadratic. Such kind of $F's$ provide model independently an Einstein frame scalar potential featuring a plateau at early times and a tail at late times, allowing the possibility for a joined explanation of the earlier and later accelerated expansion of the Universe \cite{Dioguardi:2022oqu}. However the insurgence of $\dot\zeta$ pole before the end of inflation \cite{Dioguardi:2022oqu}, makes such an idea unviable.
As shown in Section \ref{sec:FRX}, a possible solution is to upgrade to a $F(R_X)$ model.

\section{Metric vs. Palatini} \label{appendix:B}
In this appendix we give more details about the differences that arise when using the metric or Palatini formulation of gravity in the context of $F(R)$ and $F(R_X)$ models. The starting point is the following action:
\be
S \! = \!\! \int \! d^4x \sqrt{-g^J}\qty(\frac{F(R_J + \sigma X)+(1-\sigma) X}{2} - V(\phi)) , \label{eq:action:FRX:repeated}
\ee
where, again, we assumed Planck units, $\MP=1$ and a space-like metric signature. $V(\phi)$ is the inflaton scalar potential, $F$ is an arbitrary function of its argument, $X = -g^{\mu\nu}_J \partial_\mu \phi \partial_\nu \phi$ denotes the inflaton kinetic term and $R_J = g^{\mu\nu}_J R_{\mu\nu}(\Gamma)$ where $R_{\mu\nu}(\Gamma)$ is the Ricci tensor built from the connection $\Gamma_{J,\mu\nu}^\rho$. Setting
$\sigma=0,1$ allows to describe respectively the $F(R)$ and the $F(R_X)$ scenario. It is easy to check that when $\sigma=0$ action \eqref{eq:action:FRX:repeated} becomes action \eqref{eq:action:FR}, while when $\sigma=1$ action \eqref{eq:action:FRX:repeated} becomes action \eqref{eq:action:FRX}. Regardless of the gravity formulation, we can rewrite action \eqref{eq:action:FRX:repeated} in terms of the auxiliary field $\zeta$:
\be
  S_J = S_J(R_J) + S_J (\zeta,\phi) \, , \label{eq:action:zeta:J:FRX2}
\ee
where
\bea
   &&  S_J(R_J) = \int \dd^4 x \sqrt{-g}_J \left[\frac{1}{2} F'(\zeta) R_J  \right] \, , \label{eq:action:nmc} \\
   && S_J(\zeta,\phi)  = \int \dd^4 x \sqrt{-g}_J \mathcal{L}_J(\zeta,\phi)
\label{eq:action:phizeta:J}
\eea
with
\be
 \mathcal{L}_J(\zeta,\phi) = - \frac{1}{2} \left[\sigma F'(\zeta) + (1- \sigma)\right] \partial_\mu \phi \partial^\mu \phi - V(\zeta,\phi) 
\ee
and
\be 
V(\zeta,\phi) = V(\phi) - \frac{F(\zeta)- \zeta F'(\zeta)}{2} \, .
\ee
Using the EoM $\zeta = R_J + \sigma X$ into \eqref{eq:action:zeta:J:FRX2}, we restore eq. \eqref{eq:action:FRX:repeated}.
Performing the Weyl rescaling
\be
g_{\mu\nu}^E = F'(\zeta) g_{\mu\nu}^J \, , \label{eq:Weyl}
\ee
needed to move the theory to the Einstein frame, the term \eqref{eq:action:phizeta:J} becomes
\be 
S_E(\zeta,\phi) = \int d^4x \sqrt{-g^E} \mathcal{L}_E(\zeta,\phi)  \, , \label{eq:action:phizeta:E}
\ee
with
\be
\mathcal{L}_E(\zeta,\phi) = - \frac{1}{2}
\left[\sigma+\frac{(1- \sigma)}{F'(\zeta)}\right]
g^{\mu\nu}_E \partial_\mu \phi \partial_\nu \phi - U(\zeta,\phi) \, ,
\ee
and
\be 
U(\zeta,\phi) = \frac{V(\phi)}{F'(\zeta)^2} -\frac{F(\zeta)}{2 F'(\zeta)^2} + \frac{\zeta}{2 F'(\zeta)} \, , \label{eq:einstein_potential:repeat}
\ee
both in the metric and Palatini formulation (cf. the second and third terms in actions \eqref{Einstein_action:FR} and \eqref{Einstein_action}). The difference between the two types of gravity arises in the transformation of the term \eqref{eq:action:nmc}. 
Performing the rescaling \eqref{eq:Weyl} under Palatini gravity, action \eqref{eq:action:nmc} becomes simply the Einstein-Hilbert action:
\be
  S^{EH}_\text{Palatini} = \int d^4x \sqrt{-g^E} \frac{R_E}{2} \label{eq:action:EH} \, .
\ee
On the other hand, in case of metric gravity, an additional contribution is generated, giving
\be
  S^{EH}_\text{metric} = \int d^4x \sqrt{-g^E} \qty(\frac{R_E}{2} 
  - \frac{3}{2} \frac{\partial_\mu F'(\zeta) \partial^\mu F'(\zeta)}{\left(F'(\zeta)\right)^2})  \, .
  \label{eq:action:EH:metric}
\ee
The second term of eq. \eqref{eq:action:EH:metric} represents a kinetic term for $\zeta$. This is the main difference between the two gravity formulations.  Starting from a metric $F(R)$ or $F(R_X)$ theory, $\zeta$ becomes a dynamical degree of freedom after moving to the Einstein frame, implying an eventual double-field inflationary setup. On the other hand, starting from a Palatini $F(R)$ or $F(R_X)$ theory, $\zeta$ remains an auxiliary field after moving the theory to the Einstein frame, implying just a single-field inflationary setup.

\bibliographystyle{elsarticle-num} 
\bibliography{references}

\end{document}